\documentclass[acmsmall,nonacm]{acmart}
\AtBeginDocument{%
  }

\acmJournal{JACM}

\usepackage{booktabs}
\usepackage{xcolor}
\usepackage{enumitem}

\begin{document}

\title{Skills for the future software profession: beyond agentic AI !}

 \author{Abhik Roychoudhury}
 \email{abhik@nus.edu.sg}
 \orcid{0000-0002-7127-1137}
 \affiliation{%
   \institution{National University of Singapore}
   \country{Singapore}
 }

\author{Sungmin Kang}
\email{sungmin@nus.edu.sg}
\orcid{0000-0002-0298-5320}
\affiliation{%
   \institution{National University of Singapore}
   \city{Singapore}
   \country{Singapore}
 }

 \author{Baishakhi Ray}
 \email{rayb@cs.columbia.edu}
 \orcid{0000-0003-3406-5235}
 \affiliation{%
  \institution{Columbia University}
   \city{New York City}
   \country{USA}
 }

\renewcommand{\shortauthors}{Roychoudhury, Kang, Ray}

\begin{abstract}
  As coding agents are rapidly changing software engineering, a natural question is: what are the core skills needed by future software engineers? To identify where software engineering is headed and thus what skills will be needed, we summarize the results of two round-tables with researchers and industrial practitioners, held in 2026 in New York and Singapore. One key finding is that verification \& validation is increasing in importance as agents handle implementation, as highlighted by anecdotes from the events. From our observations, we identify the skills developers need in the agentic era of development, with implications for training and educating future software engineers in coming years. 
\end{abstract}

\ccsdesc[500]{Artificial Intelligence}
\ccsdesc[300]{Software and its engineering}

\keywords{AI agents, Software Engineering, Verification and Validation}

\received{2026}

\maketitle

\section{Introduction}
\label{sec:introduction}


Agentic Artificial Intelligence (AI) has changed the software engineering profession already. Agents are not only generating code, but also reviewing code, producing tests from software issues, and even helping in software design and architecture. This has created significant opportunities as well as significant disquiet among computing researchers and practitioners. Is the future of software engineering merely a collection of agents with no human in the loop ? Or does it involve agent - human dynamics in future software teams, so that we need to educate the next generation of engineers differently ? 

\begin{figure}
    \centering
    \includegraphics[width=0.8\linewidth]{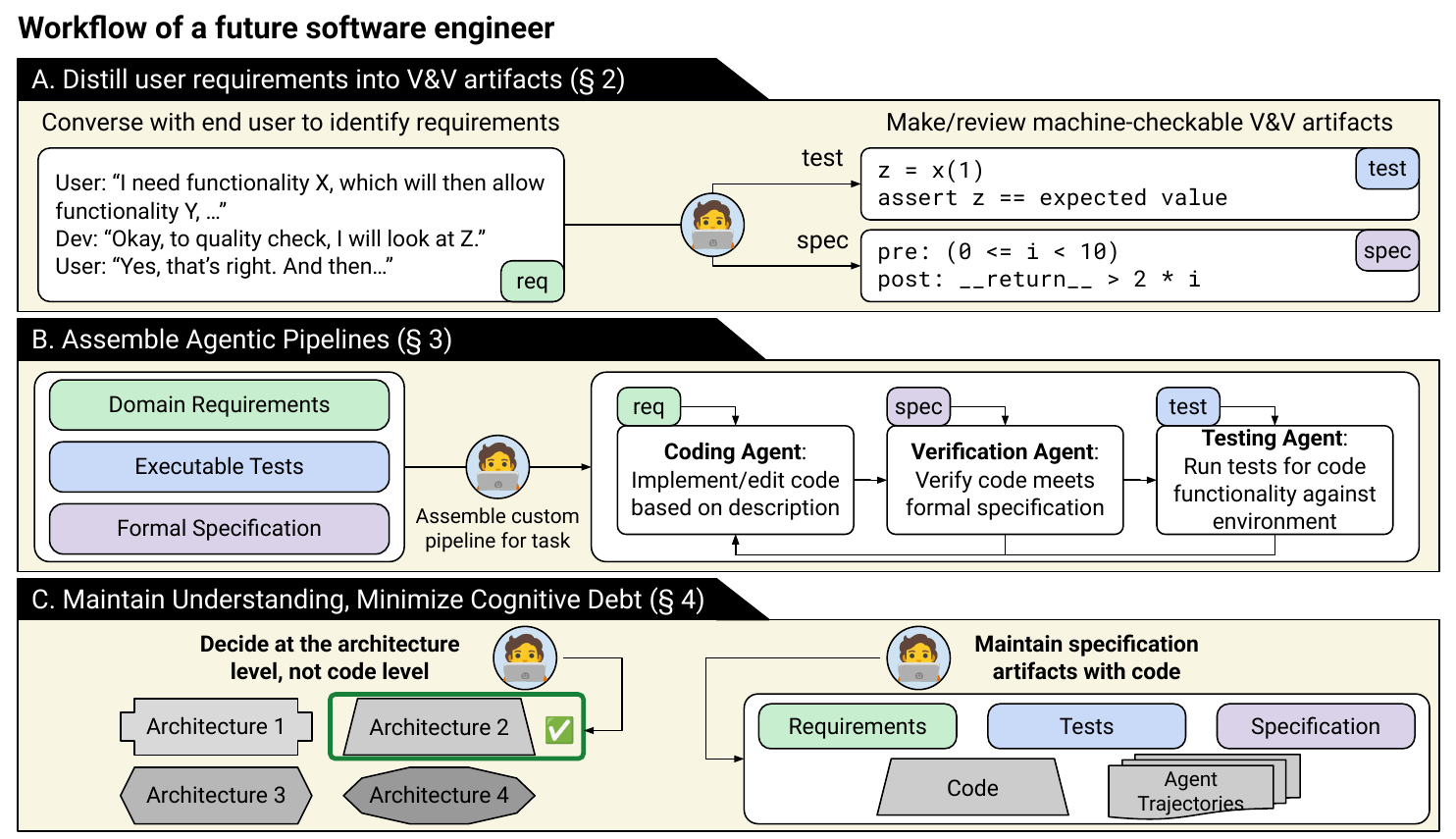}
    \caption{Workflow of a future software engineer.}
    \label{fig:workflow}
    \vspace{-5mm}
\end{figure}

To understand what would be important for future software engineers, we conducted two roundtables on the topic of Trustworthy AI for Code, inviting experts from academia and industry. The roundtables were held in January and June 2026 respectively, in Singapore and New York, each with roughly 30-40 attendees from MIT, CMU, NUS, Imperial, Columbia, Harvard, Google, Meta, Amazon, IBM, Sonar, and many others. 
Based on roundtable discussions, we propose a workflow for software engineers as depicted in Figure~\ref{fig:workflow}. It highlights the following roles for developers:


\begin{itemize}[leftmargin=*]

\item \textbf{From requirements to executable specifications.} 
Future engineers must be educated to translate user intent into machine-checkable verification and validation artifacts that guide and evaluate AI-generated implementations. Equally important, they must learn to reason about the quality of executable specifications. As AI increasingly automates specification generation, the defining human skill will be evaluating whether those specifications faithfully capture intent. 

\item \textbf{Agent orchestration.} 
Future engineers must be educated to design and reason about agentic workflows that coordinate specialized AI agents. As AI increasingly automates workflow construction, the defining human skill will be evaluating whether these workflows are robust, trustworthy, and aligned with organizational goals and constraints.

\item \textbf{Managing cognitive debt.} As AI produces an increasing fraction of software artifacts, 
future engineers must be educated to understand and manage cognitive debt in AI-assisted software development. 
They will need to preserve architectural understanding, maintain organizational knowledge, and ensure that critical design intent is not lost as systems evolve.
\end{itemize}

In the remainder of this article, we discuss each competency and its implications for educating the next generation of software engineers. Overall, the roundtables had discussion groups on
(a) role of verification and validation
(b) agentification of organisations, and 
(c) managing cognitive debt.
We summarize the discussions along these lines, with focus on education.

\section{Verification and Validation}
\label{sec:vnv}

As coding agents become increasingly capable, software verification and validation (V\&V) will play a central role in software engineering. 
Our roundtable participants observed that developers increasingly accept AI-generated code with limited scrutiny, making manual review an unreliable safeguard. Consequently, machine-checkable specifications and V\&V artifacts will become the primary basis for trust, with specification / verification preceding implementation  (Figure~\ref{fig:workflow}A).

This vision depends on the availability of high-quality specifications. Since formal specifications rarely exist at the start of a project, \emph{specification inference} becomes a critical capability. AI agents can infer executable tests and formal specifications from documentation, code, and issue reports, leaving developers to assess whether these artifacts faithfully capture user intent. Although specification inference predates agentic AI~\cite{nguyen2013semfix,kang2016apex}, it is becoming essential for trustworthy software development. As an ancedote, in the roundtable we discussed how AutoCodeRover~\cite{zhang2024autocoderover} uses inferred specifications extracted through code search to improve coding performance, an approach later commercialized in the SonarQube Remediation Agent\footnote{\url{https://www.sonarsource.com/products/sonarqube/remediation-agent/}}. Likewise, we discussed that developers increasingly rely on AI to recover code and design intent~\cite{mali26}. 
Significant research opportunities remain. Advances in LLMs have made formal verification more accessible by largely automating proof search~\cite{tao2025machine}, shifting the bottleneck from proving correctness to generating the right specifications. Verifying high-level properties often requires helper lemmas, which are localized specifications discovered through program analysis and invoked by agentic verifiers~\cite{tu2026autorocq}. While existing work focuses primarily on proof search~\cite{thompson2024rango}, roundtable participants believed that specification inference will become the more fundamental challenge for both research and practice. Inferred specifications may drift from evolving implementations, requiring consistency validation between code and specifications.

\paragraph*{Implication about education} This shift has important implications for software engineering education. Future engineers \emph{need not} become formal methods experts, but they must develop a deep understanding of formal (executable) specifications and their role in trustworthy AI-assisted development. As specification generation becomes increasingly automated, a defining human competency will be evaluating whether specifications faithfully capture user intent. This will be done by reasoning about conflicting evidence from tests, specifications, and implementations.


\section{Agentic Architecture}
\label{sec:workflows}

Once specifications and V\&V artifacts are in place, the next challenge is designing systems of collaborating AI agents to produce trustworthy software (Figure~\ref{fig:workflow}B). Rather than implementing every component themselves, future engineers will increasingly architect ecosystems of specialized agents that generate, review, verify, and refine one another's outputs. They will act as a software architect, but with autonomous agents as the primary building blocks.

Prior to arrival of agents, most software engineers were engaged in brown-field software engineering -- where a software project is evolved over time, with source code commits coming in. The source code commits correspond to program improvement such as bug fixes, feature improvement and efficiency improvement. Software education thus involves in significant part, the experiential process of inserting, managing and validating changes into the code-base. As the code {\em generation} becomes increasingly automated  - where does the balance of (remaining) activities lie? The education of these activities, particularly was the topic of discussion among round-table panelists.

\paragraph*{Implications about Education}
The answer to these questions lie in two parts. First of all, for software construction itself, the focus should turn to agent-human interaction as well as verification and validation. Secondly, the future engineers could go beyond engineering software. Instead they could focus on engineering and managing organizational workflows. Increasingly, with greater deployment of agents - organizational workflows are being "agentified". This corresponds to turning different components of an organizational workflow into automated agents - without analyzing the risks or implications on the rest of the workflow. Roundtable participants shared their experiences in agentification and specific problems that arise, such as importance of long-term management of \texttt{AGENT.md} files as a source for intent. At the same time, agents without guardrails can pose security and compliance risks; developers need to set understandable policies that cannot be evaded. Future education can focus on policy construction for risk minimization as well as the impact of agentification of a component on the rest of the organization. The engineers need to be educated about agent guardrails to achieve secure agent workflows.

\section{Managing Cognitive Debt}
\label{sec:cognitivedebt}

As agents maintain a greater proportion of software artifacts, it becomes increasingly important to manage developers' \emph{cognitive debt}~\cite{storey26}. Technical debt is a well-known concept in software engineering, arising from design and implementation compromises made to accelerate delivery. In contrast, cognitive debt emerges when developers gradually lose understanding of the intent, architecture, and rationale behind software components generated or maintained by AI agents. In the agentic era, the primary debt shifts from within the software itself (technical debt) to deficiencies in human and organizational understanding (cognitive debt). 

Agentic software development fundamentally changes how engineers interact with software. Rather than implementing functionality themselves, developers increasingly review, refine, and orchestrate the outputs of multiple AI agents. While this greatly improves productivity, it also reduces the opportunities through which engineers naturally develop a deep understanding of the systems they maintain. A developer may successfully integrate AI-generated code, tests, and documentation without fully understanding why a particular design was chosen, what assumptions were made, or which business rules must always hold. The software continues to function correctly, yet the organization gradually loses the knowledge required to confidently modify, audit, or extend it. Managing cognitive debt therefore extends well beyond preserving source code. Future software projects will increasingly include executable specifications, verification artifacts, agent workflows, prompts, tool policies, and execution traces as first-class engineering assets. These artifacts collectively capture the reasoning and intent behind AI-assisted development.
Industry participants in the roundtables discussed how they are increasingly placing a greater importance to the maintenance of specifications, prompts and trajectories along with the code generated. It was also discussed that training models with a trust mind-set is also key to generating trustworthy code. One possible avenue is to train language models with explicit knowledge of semantics~\cite{ding2024semcoder}.



\paragraph*{Implications for Education}
 The key to defining the future software engineering processes will emanate from the "authorship policy" of the code. This means that, even if agent produces code and tests, human programmers will be liable and responsible for correctness, maintainability, security, compliance of the results. This would immediately mean that for agentic coding pipelines, a commit would not be restricted to code commits. Instead, it can include artifacts such as agent instructions, tool policies, hooks and commands, skills, sandbox rules and the like. Maintaining and depositing them in a legible way can help in clearing cognitive debt of developers in future - when the opportunity arises. Multiple roundtable participants also noted the need for future developers to understand organizational knowledge and goals to help agents translate them into software. Moving forward the agents will carry the software engineering knowledge-base while humans will carry the organizational knowledge-base. Human-agent cooperation will allow their combination.



\section{Perspectives}

The defining skill of future software engineers will not be writing code. It will be translating human intent into verifiable artifacts, orchestrating intelligent agents, and maintaining trust in software systems whose implementation is increasingly automated. Computing education must evolve accordingly.
The shift to agentic software engineering is leading to non-trivial changes to the software development life cycle (SDLC), such as depositing specification in a repository (e.g. markdown files, as one roundtable participant described), discussions on whether a new specification language is needed, and how specification could be made executable to prevent specification drift. All of these changes necessitate changes in future computing curriculum.

Future engineers will need to be able to convert business requirements into checkable (executable) specifications, albeit with the assistance of AI agents. Developers will also need to compose agents and sub-agents to construct an agentic pipeline to achieve their goals. Thus, the current developer concerns about managing software scale while composing software components - may shift to future developer concerns on managing trust in agents as agentic pipelines are composed together. 
Future engineers must also acquire domain expertise in areas such as finance, healthcare, and science, where correctness depends as much on domain knowledge as on software engineering. This evolution can naturally build upon the growing ``Computing + X'' paradigm~\cite{brodley2024csx}, extending interdisciplinary programs to prepare students for AI-assisted, domain-aware software development. Ultimately, these changes redefine the role of the future software engineer. 

\section*{Acknowledgments}
We would like to thank the participants of the AI for Code Roundtables in Singapore (held on 19th January 2026) and New York (held on 3rd June 2026). This research  is supported by the National Research Foundation, Singapore, under NRF Investigatorship Program, "Agentic AI based Software of the Future: from Scale to Trust", Award ID: NRF-NRFI11-2026-0001, and National Science Foundation, USA, Award ID : NSF-CCF-2313055: "Learning Semantics of Code To Automate Software Assurance Tasks".

\bibliographystyle{ACM-Reference-Format}
\bibliography{sample-base}

@inproceedings{thompson2024rango,
  title={Rango: Adaptive retrieval-augmented proving for automated software verification},
  author={Thompson, Kyle and Saavedra, Nuno and Carrott, Pedro and Fisher, Kevin and Sanchez-Stern, Alex and Brun, Yuriy and Ferreira, Jo{\~a}o F and Lerner, Sorin and First, Emily},
  booktitle = {International Conference on Software Engineering ({ICSE})},
  year = {2025}
}

@inproceedings{nguyen2013semfix,
  title={Semfix: Program repair via semantic analysis},
  author={Nguyen, Hoang Duong Thien and Qi, Dawei and Roychoudhury, Abhik and Chandra, Satish},
  booktitle={2013 35th International Conference on Software Engineering (ICSE)},
  pages={772--781},
  year={2013},
  organization={IEEE}
}

@inproceedings{zhang2024autocoderover,
  title={Autocoderover: Autonomous program improvement},
  author={Zhang, Yuntong and Ruan, Haifeng and Fan, Zhiyu and Roychoudhury, Abhik},
  booktitle={Proceedings of the 33rd ACM SIGSOFT International Symposium on Software Testing and Analysis},
  pages={1592--1604},
  year={2024}
}

@inproceedings{tu2026autorocq,
  title={Agentic Verification of Software Systems},
  author={Haoxin Tu and Huan Zhao and Yahui Song and Mehtab Zafar and Ruijie Meng and Abhik Roychoudhury},
  booktitle={2026 ACM International Conference on the Foundations of Software Engineering (FSE)},
  year={2026},
}

@article{ding2024semcoder,
  title={Semcoder: Training code language models with comprehensive semantics reasoning},
  author={Ding, Yangruibo and Peng, Jinjun and Min, Marcus J and Kaiser, Gail and Yang, Junfeng and Ray, Baishakhi},
  journal={Advances in Neural Information Processing Systems},
  volume={37},
  pages={60275--60308},
  year={2024}
}

@article{tao2025machine,
  author  = {Tao, Terence},
  title   = {Machine-Assisted Proof},
  journal = {Notices of the American Mathematical Society},
  year    = {2025},
  volume  = {72},
  number  = {1},
  pages   = {6--13}
}

@article{brodley2024csx,
  author  = {Brodley, Carla E. and others},
  title   = {{ACM} 2023: {CS} + {X}---Challenges and Opportunities in
             Developing Interdisciplinary-Computing Curricula},
  journal = {ACM Inroads},
  year    = {2024},
  volume  = {15},
  number  = {3},
}

@article{storey26,
author = {Storey, Margaret-Anne},
title = {From Technical Debt
to Cognitive and Intent Debt -
Rethinking software health in the age of AI},
journal = {ACM Queue},
volume = {24},
number = {2},
year = {2026}
}

@inproceedings{mali26,
  author    = {Kuo, Nadine and Sergeyuk, Agnia and Chen, Valerie and Izadi, Maliheh},
  title     = {Developer Interaction Patterns with Proactive AI: A Five-Day Field Study},
  booktitle = {Proceedings of the 31st International Conference on Intelligent User Interfaces (IUI)},
  year      = {2026},
  publisher = {ACM},
  pages = {349--362}
}

@inproceedings{kang2016apex,
  title={Apex: Automated inference of error specifications for c apis},
  author={Kang, Yuan and Ray, Baishakhi and Jana, Suman},
  booktitle={Proceedings of the 31st IEEE/ACM international conference on automated software engineering},
  pages={472--482},
  year={2016}
}

\end{document}